\documentclass[11pt, a4paper, onecolumn]{article}
\usepackage{geometry}                
\usepackage[parfill]{parskip}    
\usepackage{graphicx, xcolor, fancybox}

\usepackage{boxedminipage}
\usepackage[small,bf, sc]{caption}

\usepackage{fancyhdr} 
\newcommand{\fof}{{\mbox{\tiny FoF}}}
\newcommand{\ra}{\enspace\,\,}

\newcommand{\all}{\mbox{\sc all}}
\newcommand{\lr}{\mbox{\sc lr}}

\title{\Huge The StressVaR: A New Risk Concept for Superior Fund Allocation}
\author{\Large Cyril Coste, Rapha\"el Douady, Ilija I. Zovko \\[0.2cm]
\large RiskData S.A., 6 rue de l'Amiral Coligny, 75001, Paris, France}
\date{}
\begin{document}
\maketitle




\section*{Summary}
In this paper we introduce a novel approach to risk estimation based on nonlinear factor models - the \emph{"StressVaR"} (SVaR). Developed to evaluate the risk of hedge funds, the SVaR appears to be applicable to a wide range of investments. The computation of the StressVaR is a 3 step procedure whose main components we describe in relative detail.  Its principle is to use the fairly short and sparse history of the hedge fund returns to identify relevant risk factors among a very broad set of possible risk sources.  This \emph{risk profile} is obtained by calibrating a \emph{collection} of nonlinear single-factor models as opposed to a single multi-factor model.  We then use the risk profile and the very long and rich history of the factors to asses the possible impact of known past crises on the funds, unveiling their hidden risks and so called "black swans"~\cite{Taleb07}.

In backtests using data of 1060 hedge funds we demonstrate that the SVaR has better or comparable properties than several common VaR measures - shows less VaR exceptions and, perhaps even more importantly, in case of an exception, by smaller amounts.

The ultimate test of the StressVaR however, is in its usage as a fund allocating tool. By simulating a realistic investment in a portfolio of hedge funds, we show that the portfolio constructed using the StressVaR on average outperforms both the market and the portfolios constructed using common VaR measures.

For the period from Feb.\ 2003 to June 2009, the StressVaR constructed portfolio outperforms the market by about 6\% annually, and on average the competing VaR measures by around 3\%. The performance numbers from Aug.\ 2007 to June 2009 are even more impressive.  The SVaR portfolio outperforms the market by 20\%, and the best competing measure by 4\%.


\newpage
\section{Introducing the StressVaR}
Hedge funds are a particular class of assets. They are known to have non-gaussian returns, often with a strong skew like options~\cite{Lo01,Malkiel05, Fung04}. Most of them report only on a monthly basis and often hold illiquid assets resulting in smoothed returns~\cite{Getmansky03}. With such characteristics it is very difficult to understand and estimate the possible extreme risks associated with investments in hedge funds. These risks are not easily detectable using only the series of the funds' returns and one needs to develop more sophisticated methods.

The Value-At-Risk (VaR) is a widely used and generally accepted quantitative measure of risk. A common approach to estimating the VaR is using the fund returns. However, simply fitting a Gaussian distribution to the returns in order to estimate the VaR is known to have serious problems~\cite{Adrian07}. The Gaussian assumption is well known to underestimate the likelihood of large returns and consequently extreme risks. The Cornish-Fisher VaR~\cite{Cornish37} takes a step in the right direction by a better modeling of the fat tails, taking into account the third and fourth moments of the return distribution. Different models of the returns distributions lead to other statistical measures based on historical returns, for example the GARCH based VaR, Extreme Value Theory, Omega ratio, and nonstationary models~\cite{Angelidisa04,Giot04,Hangchan07}.

However, all of the above statistical measures of the VaR suffer from similar problems which are rooted in the fact that they are estimated from historical returns. The most numerous returns in the historical returns are small returns. Therefore, the largest information input in the historical based measures comes from small returns. This is quite in an opposite spirit of what a good risk measure needs to be.

In order to assess the risk of a hedge fund, the large returns or outliers carry the most amount of information. It is in the outliers that the signal for potential problems lies, not in the business-as-usual returns. This is also the reason why robust methods for risk estimation, which try to limit the influence of outliers, are in fact very much against the principles of risk estimation. 

But in fact the problem goes even deeper. The fund's risk, or the possibility of a future loss, is very often simply not contained in the historical return series of the fund. Somehow one needs to predict the possible behavior of the fund in scenarios that \emph{have not yet happened} during the fund's existence.  

The StressVaR we propose here goes in this direction in spite of being partially based on historical fund returns.  The risk information in the StressVaR comes both from the dynamical relation of the fund and a large number of factors, as well as from the behavior of the factors themselves, for which we can have a much better understanding of their risks.  By modeling the fund's performance using factors for which long term history is available we benefit from the richness of crisis information contained in the factor history.

For example, even though a fund has not yet lived through its greatest crisis, one can predict the risk of the fund by extrapolating the fund's behavior into the past using factors. In other words, a fund can be subject to a risk that hasn't occurred in its history and hence is not present in the historical returns. Had the fund been incepted earlier, the risks would show in the fund returns. Loosely speaking, it is these kinds of risks that Taleb calls Black swans\footnote{Strictly speaking Taleb's black swan is an "unknown unknown" -- an unpredictable event one did not even conceptualize as possible. We are referring to the "known unknowns" -- unpredictable events that are hidden in the history of a particular asset, but can be conceptualized as possible if one looks outside of the fund's history.}~\cite{Taleb07}.

The commonly used alternative approach -- erroneously considered as "conservative" -- is to base the risk analysis on the fund holding's. This approach suffers three major limitations:
\begin{itemize}
\item Position data are only partially available, delayed and often contain errors.
\item An accurate joint distribution of the holdings returns is necessary. Such a distribution is hard to estimate and often very rough approximations are made to handle the large dimensionality of the problem. These approximations -- correlations, liquidity, etc. -- are often invalid under extreme events. Moreover, some of the holdings may have a short history, or be hard to value, leading again to an inaccuracy in the distribution of returns.
\item Hedge funds positions experience a high turnover. Delayed and past positions may be unrelated to future ones. Moreover, any non-zero correlation between the position changes and returns of the corresponding assets produce nonlinearities -- exactly in the way how dynamic hedging allows replicating an option pay-off.
\end{itemize}

The StressVaR can be compared to a factor-based VaR estimate in the sense that is uses factor models. However, its strength resides in the modeling of nonlinearities and the capability to analyse a very large number of potential risk factors. Regarding the information needed, its computation only requires the fund's returns and general market history -- not specific to the analyzed funds. In this regard, it allows a full analysis of portfolios of hedge funds, regardless of the knowledge of detailed holdings and positions.

Estimating the StressVaR is a 3 steps method that starts by selecting a large universe of factors. It is important that the universe is large enough to capture all possible risk sources for a hedge fund in question. Following that, the estimation of the StressVaR consists of the following steps:
\begin{enumerate}
\item \textbf{Factor scoring.} For each of the factors in the universe one estimates a dynamic nonlinear model of the fund against the factor obtaining a goodness of fit measure, for example the p-value. The factor p-values are then used to score and rank the factors in terms of their explanatory power for the fund returns. Typically one focuses on a certain number of top explanatory factors whose p-values pass a threshold denoting that they are not representing spurious relationships.
\item \textbf{Estimating factor risks.} For each of the top factors, using the calibrated model, one predicts the fund returns for all possible factor returns from the 1st to the 99th quantile of the long term factor return distribution.
\begin{itemize}
\item [a)] Estimation of the factor distribution.  Estimating the factor distribution is a separate and challenging issue in its own right, and is completely independent from the fund return prediction.  For example, one can use an ARCH type model for estimation, however it is important to use as long a factor history as possible in order to account for known past crises.  In addition, the estimate of the factor distribution can also be aided by pure economic studies.
\item [b)] Predicting the fund returns.  Following the estimation of the factor distribution, one predicts possible fund losses for a given percentage $q$ of factor returns. For example for $q=98\%$ one would estimate the fund returns from the 1st to the 99th quantile of the factor distribution, leading to the estimation of the 98\% VaR.\footnote{In fact 99\% VaR if the relation between the fund and factor is monotonously increasing or decreasing.}  By taking the min over the predicted fund returns one obtains the largest possible fund loss corresponding to $q\%$ of the factor returns.
\end{itemize}
\item \textbf{Estimating the StressVaR.} Ultimately, the StressVaR is estimated as the maximum predicted loss across all top selected factors to which the specific risk, estimated in parallel from the model residuals, is added. Arguably, this estimate for the VaR overestimates the real VaR.  However, as we will show, in practice this approach ultimately leads to better investment decisions. 
\end{enumerate}

It is important to understand the interplay of different time horizons in the estimation of the StressVaR. The dynamical relation between the fund and the factors is estimated using the shortest\footnote{One of course needs to use a sufficient amount of data for a reliable estimate, but adding more points than minimally necessary is not advantageous.} and most recent time interval. This is important in order to reactively capture changing fund strategies and factor exposures. On the other hand, the factor returns used to predict possible fund losses need be from the complete factor history, using as much data as possible. Intuitively, by using the complete factor history, we are augmenting the information contained in the history of the fund, with the information contained in the factor history. The histories of the factors are typically much longer than the lifetime of the fund and contain information of many more crisis periods bearing risk.


From a practical and estimating perspective the StressVaR approach has several benefits. Since factors are tested against the fund one-by-one, issues related to over-fitting due to a small number of data points are limited.  Furthermore, colinearity among factors does not influence the stability of estimation. Hence, the StressVaR allows to work with a virtually unlimited number of factors.

In theory, the StressVaR, computed as above, underestimates the real VaR because the effects of possible fat tails of the fund specific risk with respect to selected factors - the residuals of the nonlinear model - are ignored. However, this underestimation is limited, provided the factor set covers a complete enough range of risk sources. Indeed, in most cases the fat tails of the fund are caused by one of the factors.  An extreme event of the fund is linked to an extreme event of one of the selected factors. In this case, the residual risk is negligible with respect to the explained risk since the correlation between the fund and the factor becomes close to 1.

In practice one can on the contrary argue that the StressVaR will tend to overestimate the risk. However, what ultimately is important for risk management is not the number of time that a fund is above its q-th quantile, but that it leads to good investment selection decisions. In other words, that the StressVaR was a good estimator of the future move of the funds is perhaps less important than that it really distinguishes between risky and less risky funds. A good risk measure should be able to split risks into "good" and "bad" and become a major tool in the investment process. A good risk-measure in this case will be a tool for portfolio construction by rating the funds according to their risks. It should be reliable in the sense that the impact is driven by the joint behavior of that fund with the portfolio under extreme conditions and not only on normal conditions.

{\bf Liquidity risk} \\
The nonlinear and dynamic requirement of the StressVaR are warranted in extreme market conditions where market movements induce massive orders, thus creating \emph{liquidity traps} in which asset prices experience a strong slippage. The phenomenon is responsible for so-called \emph{correlation breaks} that destroy the diversification of apparently well balanced portfolios.  By correctly addressing the nonlinear relation between market factors and fund returns, the StressVaR is the first risk measure that provides a true quantitative estimate of the impact of the vanishing market liquidity under an extreme event. Illiquid assets, the prices of which are marginally correlated to markets under normal market conditions, but materially react to large moves, are also better handled by nonlinear models involved in the StressVaR than by other techniques.

{\bf Credit risk} \\
Another benefit of nonlinear factor models is in their ability to capture credit risk.  It is common practice to consider that credit risk can be identified with default risk.  But in fact, credit risk can not be reduced to pure default risk and must cover a much broader range of market events and in particular the possible explosion of credit spreads.  A good modeling of extreme market events will therefore also capture credit risk.  For example, the risk resulting from a defaulting security is typically summarized in the price of the distressed security.  Generally speaking, the relation of credit spreads with other market factors, equity prices, implied volatility, etc., is known to be extremely nonlinear and subject to threshold effects.  Nonlinear factor models are again the most appropriate way to capture, not the default probability \emph{per se}, but the real price impact of credit events.
In the particular case of CDS where the price of the credit derivative security is related to a default event, one observes that the default event always occurs jointly with the sharp drop of the stock price.  Again we see that default risk is associated with some sort of extreme market risk.  

The organization of the rest of paper is the following. In the first part we define the StressVaR in more detail than was given in the introduction. In the second part we compare the StressVaR against two traditional VaR estimators.  This is done by backtesting the number of exceptions on a broad sample of hedge funds reporting to the HFR database.   In the third part we investigate the SVaR as a fund selection and portfolio allocation tool.  We take the role of an investor in hedge funds and show that a portfolio constructed using the SVaR outperforms portfolios constructed using standard risk measures.  The StressVaR portfolio produces superior returns by eliminating the "bad risk" while keeping the "good risk".


\section{Calculating the StressVaR}
Factor selection and ranking is the most important element in the estimation of the StressVaR. The estimation methodology needs to be sensitive to a great variety of fund-factor dependencies but at the same time minimize spurious factor selection. The main objective is to specify and estimate the function of the conditional expectation of the fund given the factor $\Psi=E(Y|X)$. There are several important elements that this function needs to capture.
\begin{itemize}
\item \textbf{Nonlinearities.} It is well known that due to their active management and investments in derivatives the returns of hedge funds can be quite nonlinear~\cite{Fung97}\footnote{Internal Riskdata study jointly with E.Derman has shown that around 75\% of hedge funds are better modeled by nonlinear functions.}.
\item \textbf{Dynamic relations.} Due to the delayed nature of fund return reporting, the impact of a factor on a fund can be delayed by a few weeks or even in some cases months. The function $\Psi$ must therefore allow for lagged dependencies of the fund and factor.
\item \textbf{Autocorrelations.} The illiquid assets some hedge funds hold are typically valued in a conservative fashion, for example by only valuing the coupon payments of a bond investment. This procedure mechanically induces autocorrelations in the fund returns. The function $\Psi$ must allow for AR terms.
\item \textbf{Cointegration.} In some cases, the relation between the fund and the factor is best captured by looking at the levels rather than the returns. This is the case when the fund and the factor have a common driver and move together, but otherwise the fluctuations around this common trend are independent. The function $\Psi$ in this case needs to allow for cointegrating relations between the fund and the factor~\cite{Maddala99}.
\end{itemize}

Taking into account these properties it is useful to separate $\Psi$ into an autoregressive (AR) part that captures the autocorrelations in fund returns and the remaining part that is a function of the factor returns:
\begin{equation}
\Psi=\mbox{AR}(Y)+\Phi(X).
\end{equation}
The relevance of a factor can now be tested by comparing the full model for $\Psi$ against the AR model:
\begin{eqnarray}
& H_1:& Y_t=\mbox{AR}(Y)+\Phi(X)\nonumber\\
& H_0:& Y_t=\mbox{AR}(Y).
\end{eqnarray}
A simple way to test this is using an F-test based on the residuals of the two regressions:
\begin{equation}
F=\frac{n-k}{q}\;\frac{R_1^2-R_0^2}{1-R_1^2}
\end{equation}
where $n-k$ is the number of degrees of freedom in $H_1$ (number of points $n$ minus the number of parameters plus the constant) and $q$ is the number of extra parameters present in $H_1$ and not in $H_0$. $R_0^2$ and $R_1^2$ are the standard $R^2$ measures for the two models. The factor selection can then simply be based on whether the p-value associated with the F-test is below a certain preset threshold $\theta$.

However, one needs to keep in mind that the overall p-value of the selected set is not the same as the threshold $\theta$. If one tests $N$ different factors, \emph{assuming tests are independent}, the probability that at least one factor be spuriously selected can be as large as $\theta_{N}=1-(1-\theta)^{N}$. When -- as it happens in practice -- factors are not independent, $N$ should be replaced by some kind of an "efficient number of factors", based on an entropy measure. The number of factors should therefore be as small as possible, while still large enough to capture all risk sources. One of the known techniques to remedy this "power vs. size" problem is to have an adaptive threshold $\theta$ depending on each tested fund, and more precisely, on the range of p-values across the various factors. When a fund selects at least one factor with a very low p-value, then factors with a relatively high p-value should be ignored in order to avoid spurious selections. But when none of the factors display a sufficiently significant relation with the fund, then the threshold should be relaxed in order to avoid missing important risks. Consequently the threshold $\theta$ will be chosen as an increasing function of the smallest obtained p-value across the whole factor set.

Computing the p-value is less simple than it seems. Classical computations, such as the F-test, the Likelihood ratio or the Wald statistics have hidden assumptions about Gaussian residuals, non endogeneity, etc. In practice, the computation of the StressVaR can be significantly improved by properly taking into account the distribution of residuals and the particularities of \emph{small samples}~\cite{Ullah04, Seber89}. 

\subsection*{Prediction function $\Psi$}
The central element of the factor selection process is the proper specification of the prediction function describing the response of the fund
to the factor. As noted before, this function needs to allow for several types of joint behaviours. It is however not easy to balance the sophistication of the model with the stability of estimates given the very limited number of data points available for hedge funds. We have tested various models, including cointegrating models, but with mixed results primarily because of the data availability issue. In the end we settle for a minimum model sufficient to capture the main features of active management present in hedge funds. This model is in terms of fund and factor returns only and is of the form
\begin{equation}
dY_t=\mbox{AR}(dY) + \Phi(dX).
\end{equation}
The dynamic nature of a possible factor relation is incorporated by including a certain number of lags in the specification:
\begin{equation}
\Phi(dX)=\Phi_0(dX_t)+\Phi_1(dX_{t-1})+...
\end{equation}
The nonlinearities are captured through polynomials:
\begin{equation}
\Phi_j(dX_{t-j})=\sum_i  \alpha_i  \;  \phi_i(dX_{t-j}),\;\forall j
\end{equation}
where $\phi_i(\cdot)$ are some appropriately chosen base functions. All in all, the full model for the fund returns can be written
\begin{equation}
dY_t=\mbox{AR}(dY)   + \sum_i \alpha_{0,i} \; \phi_{0,i}(dX_t) + \sum_ i\alpha_{1,i}  \;  \phi_{1,i}(dX_{t-1}) + ...
\end{equation}
The first term (AR) captures the potential smoothing of fund returns, the second captures the instantaneous nonlinear factor response, while each successive term captures nonlinear delayed factor responses due to for example liquidity.

We made the specification of the model to be data driven - we do not choose in advance the number of lags or nonlinear terms, but rather let the model be specified as best dictated by the data. Using an OLS estimator, we introduce penalization terms for large and nonlinear models so that the estimated model is the least complicated one possible for the level of residual sum of squares.

\section{VaR backtesting}
An extensive test of the StressVaR is done using the Hedge Fund Research (HFR) database  from Feb.\ 2003 to June 2009.  We removed index funds and FoFs and then selected the funds based on the condition that they were continuously reporting on a monthly basis to the database for the entire period from Feb.\ 2000 to Feb.\ 2008.  We keep only one fund per manager (i.e., remove alternative currency versions of funds or off-shore versions).  This leaves a selection of 1060 funds in Feb 2008.  In 2008 and 2009 about 30\% of these funds stopped reporting to the database.  To deal with this, we assumed a 30\% loss of the fund in the month when the fund stopped reporting and dropped it from subsequent pool of funds.  In June 2009 there are only 764 funds remaining.

The factor set used in the estimation of the StressVaR contains 172 factors covering world-wide equities, interest rates, volatilities, credit spreads, as well as some special factors such as market convergence, correlations, etc., and goes back to 1987. 

We set the VaR percentile $q=98\%$ and compare the performance of the StressVaR 98\% to the classical VaR 99\% estimate based on the Gaussian assumption:
\begin{equation}
\mbox{GVaR}=z \cdot \sigma
\end{equation}
where $\sigma$ is the historical standard deviation and $z=2.33$ is the Gaussian 99th percentile in units of $\sigma$; as well as to the Cornish-Fisher VaR 99\% which attempts to better estimate the VaR by using the skew and kurtosis:
\begin{equation}
\mbox{CFVaR}= \sigma \{z  +\frac{1}{6}(z^{2}-1)s +\frac{1}{24}(z^{3}-3z)k- \frac{1}{36}(2z^{3}-5z)s^{2}\,\}
\end{equation}
where $s$ and $k$ are respectively the skew and excess kurtosis~\cite{Cornish37}.

\begin{table*}[tbh]
\centering
\fcolorbox{gray!20}{gray!6}{
\begin{minipage}{0.9 \textwidth}
\centering
\begin{tabular}{l r r r r r}
VaR 99 estimator	&  1$\cdot$VaR &  2$\cdot$VaR & 3$\cdot$VaR & Avg. exc. & Med. exc.	\\
\hline
Gaussian VaR		&2.11\%	&0.18\%	&0.06\%	&1.43	&1.26 \\
Cornish-Fisher VaR 	&0.82\%	&0.13\%	&0.08\%	&2.58	&1.27 \\
StressVaR	 		&1.15\%	&0.10\%	&0.03\%	&1.41	&1.21 \\
\hline 
\end{tabular}
\caption{Comparison of the exception rate and exception magnitude for the three tested VaR measures.  The leftmost 3 columns show the percentage of VaR exceptions for one, two and three times the VaR threshold.  The rightmost two columns display the magnitude of VaR exceptions in units of the VaR: the average and the median.  The reason for the SVaR 1.15\% VaR exception rate is that, strictly speaking, one should test the SVaR with $q=99.5\%$ in order to reach a 1\% exceptions rate.  However, since for most factors the relation between the fund and factor is typically monotonous, the SVaR 98\%, that is with $q=99\%$, is still close to 1\% exception rate.  What is perhaps more important is to notice that the SVaR has significantly lower 2$\times$ and 3$\times$VaR exception rates and at the same time, when a VaR exception does occur, it is exceeded by about 40\% on average compared to the 160\% for the Cornish-Fisher VaR.}
\label{VaRexceptions}
\end{minipage}
}
\end{table*}

A standard procedure in VaR backtesting is counting the number of VaR exceptions - counting the number of times a fund return is larger than the predicted VaR. While we believe that a proper test of the usefulness of a VaR measure is ultimately its performance in asset allocation, we summarize in Table~\ref{VaRexceptions} the estimated frequency of exceptions for various VaR levels.

In the leftmost three columns of the table we list the frequency of returns larger than 1 times the VaR, 2 times and 3 times the VaR.  In the rightmost two columns we also table the average and median magnitude of the VaR exception,  in units of the VaR. This illustrates the amount of potential losses a manager will suffer if the VaR estimate is broken.  Indeed, by requiring an economic capital equal to 3 times VaR the regulator assumes that this threshold is not supposed to be surpassed.  Counting such exceptions is therefore mandatory.  

The StressVaR has less exceptions than the other two measures, but what is more important is that the average and median exception sizes are about 40\% and 20\% of the VaR. In this respect, all three VaR measures perform well. One exception is the mean size of the Cornish-Fisher exception which shows that the measure is not too reliable in some situations.

While the usual number of exceptions measure takes in account only the tails of the distribution, the method of estimating the magnitude of VaR exceptions focuses in contrast on the entire distribution of fund returns.  We can accept the notion that business-as-usual is when fund returns are drawn from a Normal distribution. However, it is well known that often this is not the case -- large fund returns are more common than would be expected were they drawn from a Gaussian. It can be argued that a good risk measure should provide a normalization of actual fund returns to be more closer to a Gaussian. This is the notion around which we construct the last method of comparison of the StressVaR and the standard VaR. We show that fund returns divided by the VaR are closer in distribution to a Gaussian if we estimate the VaR using the StressVaR instead of the standard VaR.

\section{Allocation using the StressVaR}
As noted earlier, the number of exceptions may not be a particularly useful criterion for a risk measure. A good risk measure needs to prove itself capable in predictively separating risky and less risky funds. Therefore, the performance of a portfolio constructed using the risk measure is the criterion we believe is ultimately useful in discriminating among risk measures.

In this section we present the results of a backtesting study using the risk measures as an allocation tool. We take the role of a fund-of-funds investor: The investor rebalances a portfolio of hedge funds each 3 months, choosing a number of funds from the universe to invest in, together with the corresponding weights. The choice of hedge funds and their weights is done solely using one of the risk measures, i.e., based only on the quantitative information contained in the \emph{past} fund and factor returns. Once the selection of funds and their allocation is decided, we calculate the realized returns of the portfolio in the next three months. By repeating this procedure each three months and cumulating the returns we obtain the performance such a strategy would have yielded in the tested period.

The allocation decision starts by ranking all the 1060 funds in the fund universe according to their predicted risk by using one of the aforementioned risk measures. The investor then selects, say, one quarter of the least risky funds and assigns the weights inversely proportional to the risk measure $w_i = 1/\rho_i$. In this way one obtains an equal risk portfolio of the least risky funds. In order to avoid possible situations with excessive weight assigned to a perceived low risk fund, we cap the weight of a fund in the portfolio to 10\% of the total portfolio. The weights are normalized $1 = \sum w_i$, leading to:
\begin{equation}
w_i = \frac{1}{\rho_i \cdot \sum_j 1/\rho_j}.
\end{equation}
The next period return of the FoF is then:
\begin{equation}
r_\fof =\sum_i w_i \cdot  r_i = \frac{\sum_{i \in \lr} 1/\rho_i}{\sum_{j \in \all} 1/\rho_j}  \cdot  r_i
\end{equation}
where $\rho_i$ is the risk predicted in the previous period and $r_i$ the return of fund $i$ in the current period.  $\all$ and $\lr$ denote the set of all funds and the set of selected low risk funds.

Since the allocation we are investing in is by construction the quartile of least risky funds, the performance of the portfolio is also expected to be smaller - less risky funds on average will also bring smaller returns. For this reason we leverage our investment to match the global market risk. The risk adjusted portfolio return is:
\begin{equation}
\tilde{r}_\fof=(1-\lambda)\;  r_f+\lambda  \;r_\fof
\end{equation}
where the leverage factor $\lambda$ is equal to the ratio of the global risk $\Omega$ and the portfolio risk $\Omega_\fof$, and for realism capped at 3:
\begin{equation}
\lambda=\min(3, \frac{\Omega}{\Omega_\fof})
\end{equation}
and $r_f$ is the risk-free rate such as the Libor. The global risk $\Omega$ and portfolio risk $\Omega_\fof$ are averages of fund risk measures respectively for all the funds in the universe and for the funds in the portfolio (i.e. assuming no risk diversification) so that leverage ratio $\Omega / \Omega_\fof$ is
\begin{equation}
\Omega / \Omega_\fof = \frac{4 \sum_{i \in \lr} 1/\rho_i}{\sum_{j \in \all} 1/\rho_j}.
\end{equation}

\subsection*{Performance}
\begin{figure*}[tbh] 
\centering
\fcolorbox{gray!20}{gray!6}{
\begin{minipage}{0.96 \textwidth}
\includegraphics[width=\textwidth]{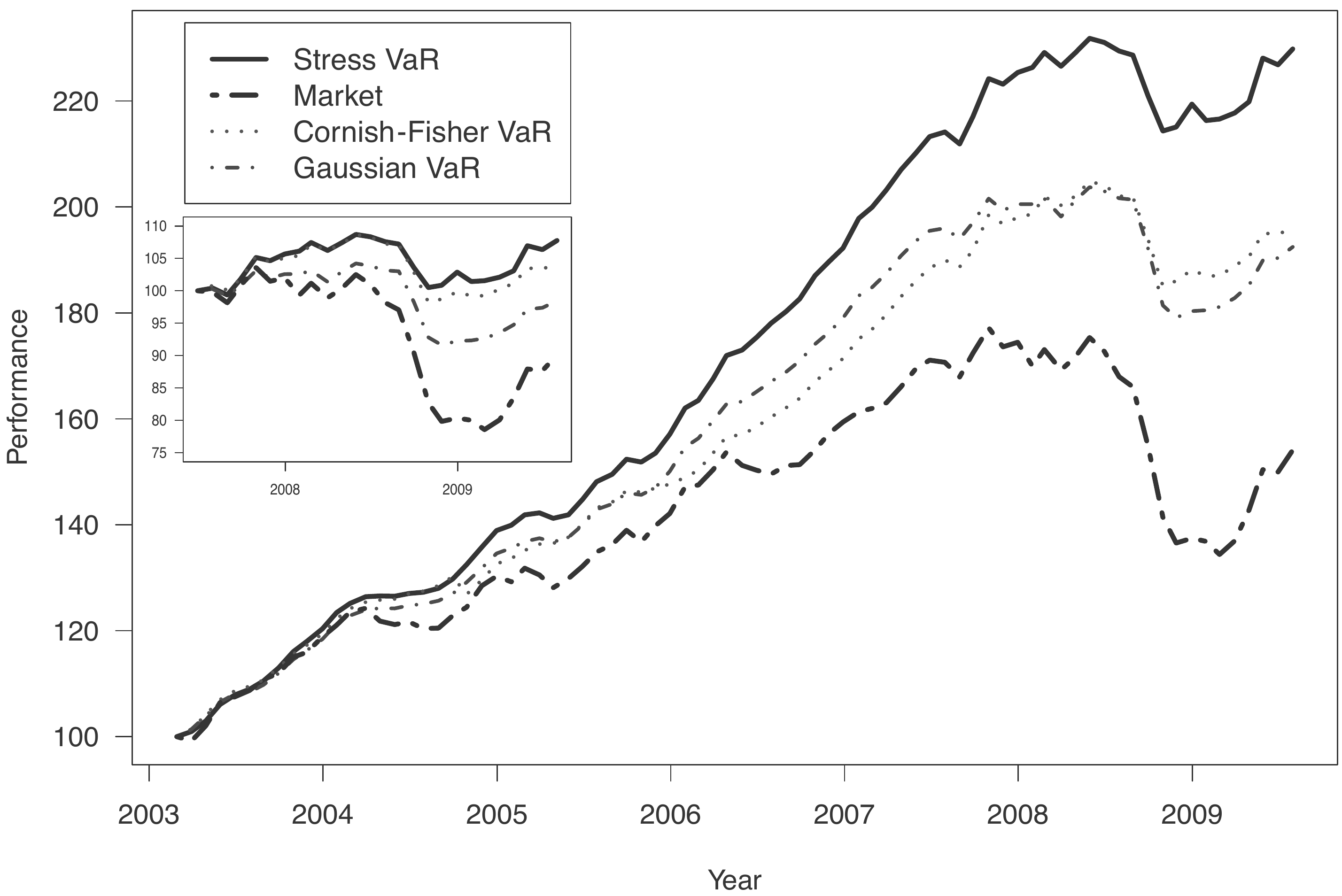}
\caption{Comparison of the performance of portfolios allocated using the tested risk measures. We see a constant outperformance of the StressVaR allocated portfolio over both the market benchmark and the portfolios allocated with the other VaR measures.  In six and a half years the gain of the SVaR allocated portfolio are about 46\% or 6\% a year.  In the inset of the figure we normalize all the portfolios to value 100 in Aug.\ 2007 in order to compare the performance only for the difficult years of 2007, 2008 and 2009. In these years the SVaR portfolio is again the best performing one and outperforms the market by about 20\%.}
\label{fig:performance}
\end{minipage}
}
\end{figure*}

The usual performance indicators for the allocation strategies using the three risk measures are shown in Table~\ref{perfs}. We can see that in terms of performance the StressVaR allocated portfolio outperforms both the market and the other two portfolios  with a comparable annual volatility.  The Sharpe ratio is correspondingly higher than for the other portfolios.  All other usual performance measures show the superior returns of the StressVaR over the benchmarks.  The performances of the portfolios are shown in Figure~\ref{fig:performance} where we can observe again a steady over-performance of the StressVaR portfolio over the other. Again, what is interesting to note is that none of the other three VaR measures manage to outperform the market benchmark.
\begin{table*}[t]
\centering
\fcolorbox{gray!20}{gray!6}{
\begin{minipage}{0.9 \textwidth}
\centering
\begin{tabular}{l  r r r r}
					&Market	&SVaR	&GVaR	&CFVaR	\\
\hline
Sharpe ratio			& 0.8\ra	&2.9\ra	&2.1\ra	&2.3\ra	\\
Annual performance		& 6.2\%	&13.0\%	&10.0\%	&10.2\%	\\
Annual volatility		& 7.8\%	&4.5\%	&4.8\%	&4.5\%	\\
Max drawdown			&-27.6\%	&-7.8\%	&-12.9\%	&-10.1\%	\\
\% positive months		& 68\%	& 83\%	& 85\%	&85\%	\\
Max time to recovery (months)	&6\ra		&5\ra		&6\ra		&5\ra		\\
\hline
\end{tabular}
\caption{Comparison of performance measures for three simulated FoF investment strategies. Each strategy is constructed by allocation decisions based on one of the three risk measures: Gaussian VaR (GVaR), Cornish-Fisher VaR (CFVaR) , and StressVaR (SVar).  For comparison the column named \emph{Market} is the performance of a portfolio equally weighted over all the hedge funds in the investment universe.  One can see that all three portfolios outperform the market.  However, the SVaR portfolio has both the highest annual performance and lowest volatility resulting in a Sharpe ratio of 2.9.  Perhaps more importantly, the max-drawdown of the SVaR portfolio is substantially lower the for the other two portfolios.}
\label{perfs}
\end{minipage}
}
\end{table*}

\begin{figure*}[t!b] 
\fcolorbox{gray!20}{gray!6}{
\begin{minipage}{0.98 \textwidth}
\centering
\includegraphics[width=\textwidth]{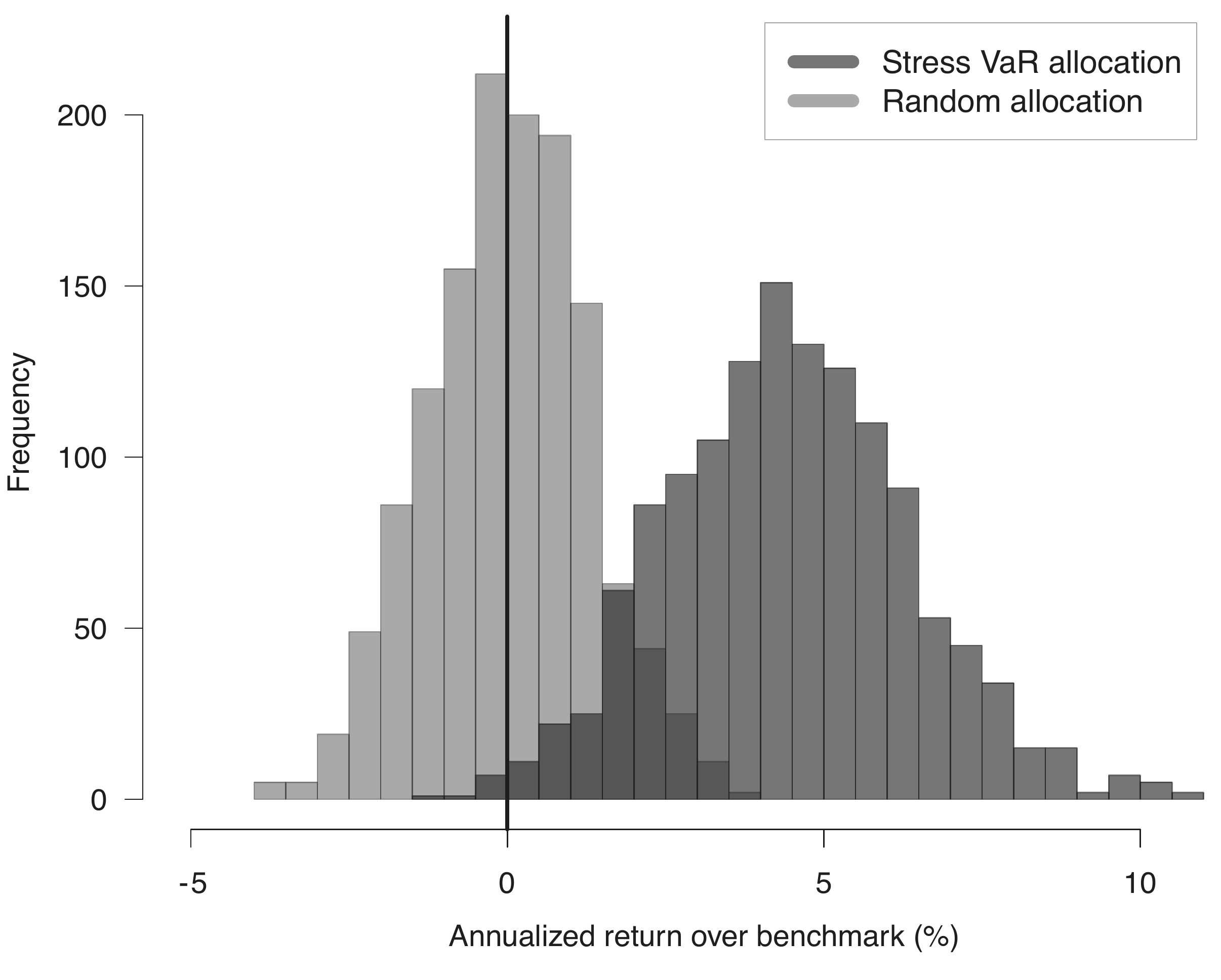}
\caption{The density estimate of annualized excess returns over the market benchmark (equally weighted index) for portfolios allocated using the StressVaR (in dark gray).  The light gray is the density of randomly allocated portfolios of equal size.  The estimates are a result of 5000 simulations where an investor makes portfolio allocations in 10 hedge funds based on the SVaR from a pool of 100 randomly chosen funds from the HFR database. More details are in the text.  The average excess return of  randomly allocated portfolios (light) is expectedly equal to zero.  However, StressVaR allocated portfolios (dark) outperform the market on average by about 5\%.  Almost never (probability is less then 0.5\%) does a SVaR allocated portfolio underperform the market (the area of the blue histogram left of zero).
}
\label{perfHist}
\end{minipage}
}
\end{figure*}

One could argue that the fund universe of 1060 funds is too large for a single investor to follow\footnote{Though there exist tools that allow for rapid screening of a large number of hedge funds by multiple criteria.} or that rebalancing a portfolio of 250 hedge funds is not realistic. With this in mind we repeat the allocation analysis in a different manner. From the 1060 funds we randomly select 100. This is now the fund universe for an investor. The investor selects 10 out of these 100 in the same way as before and rebalances quarterly. Since this simulation setup is more noisy than when using more funds, we repeat the simulation 5000 times, simulating 5000 different investors. In Figure~\ref{perfHist} we show the distribution of the annualized performance using the StressVaR versus the performance of a market portfolio. It is evident that portfolios allocated using the StressVaR outperform the market portfolio. Almost never does a strategy using the StressVaR loose money, and on average it earns 5.5\% above the market.  Using the other risk measures for allocation leads also to excess returns, but not as high as when using the StressVaR. The Gaussian VaR earns on average 2.9\% above the market and makes losses compared to the index in about 4\% of cases. The Cornish-Fisher VaR earns 4.4\% annually on average and looses money in about 1\% of cases.

The choice of quarterly rebalancing was done as a tradeoff between short rebalancing intervals, where the risk allocation produces best results, and long rebalancing intervals which are more realistic.\footnote{The authors would like to thank Gumersindo Oliveros for useful remarks and suggestions.} We have tested rebalancing intervals up to 6 months at which we still find a very significant over performance of the SVaR measure over the market and the other measures. This gives us confidence that a portfolio constructed with the SVaR really is better performing than the other benchmarks at multiple time horizons.

\subsection*{Downside protection}
\begin{figure*}[p] 
\fcolorbox{gray!20}{gray!6}{
\begin{minipage}{0.98 \textwidth}
\centering
\includegraphics[width=\textwidth]{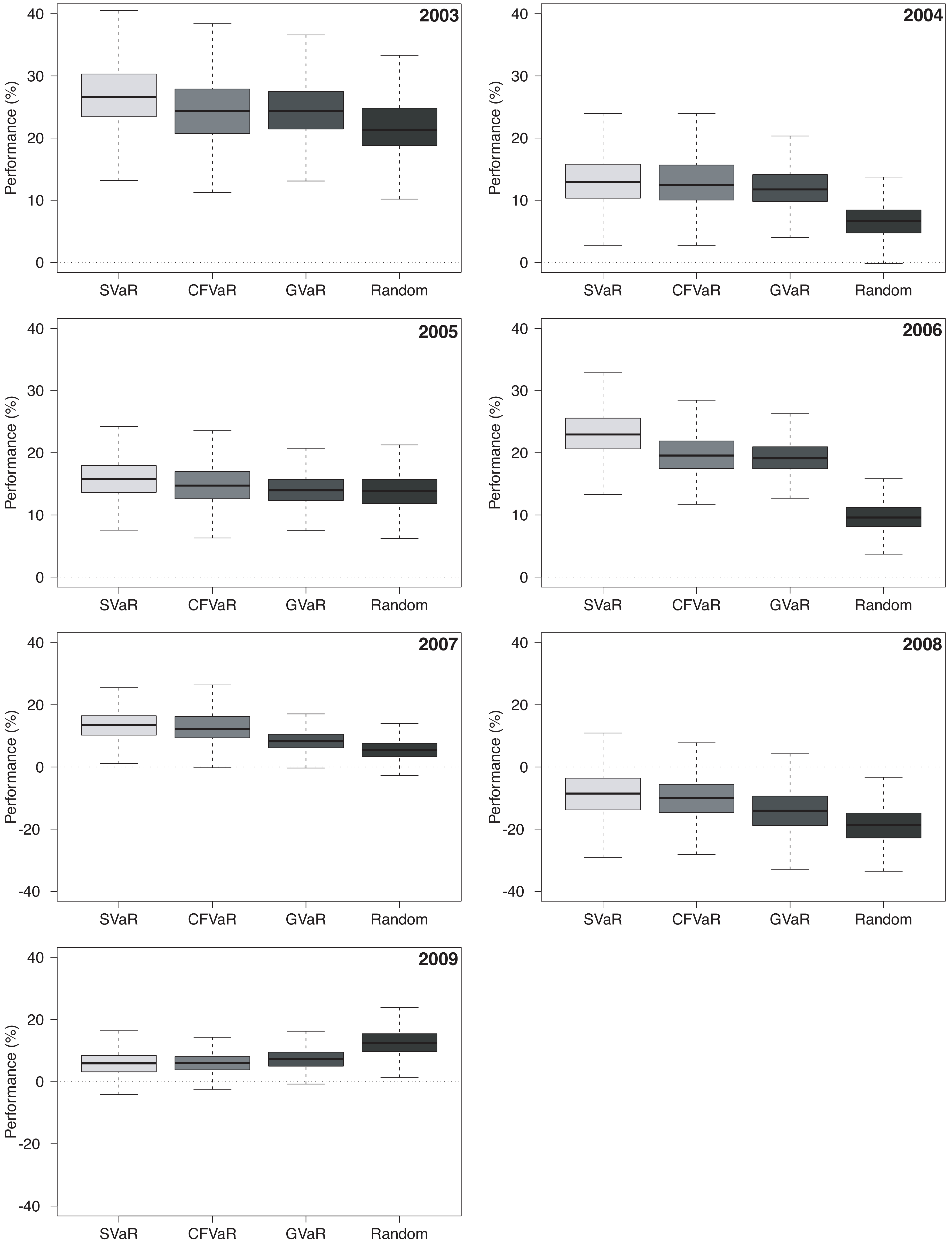}
\vspace{-0.5cm}
\caption{Boxplot comparison of yearly performance for 5000 simulated FoF portfolios allocated using the three risk measures.  Each subplot is showing the performance in the respective year.  The random portfolio is constructed from an equally weighted random selection of funds.  We see that while all risk measures do provide an improvement over the random allocation, only the SVaR consistently provides downside protection.}
\end{minipage}
}
\label{yearlyPerfBox}
\end{figure*}

In Figure~3
we repeated the portfolio allocation simulation for each of the 7 analysed years separately. We show the box-plots of the yearly returns for the simulations for the various VaR measures as well as for the random allocation.  Using any VaR measure is already an improvement, however in all cases the SVaR allocation is producing on average highest returns. Even more importantly, the SVaR provides a substantial improvement over other risk measures in terms of down side protection.  Notice that only the SVaR rebalanced portfolios capture the positive market returns of the years up to 2007, while at the same time providing downward protection in the crisis 2007 and 2008 years.  At the end of the analysed dataset, by the month of June 2009, the year 2009 is experiencing a partial recovery from the crashes of the previous years which none of the 3 portfolios are capturing.  However, all 3 are recording healthy profits.  The performance of the StressVaR is exactly what a risk measure should do: provide downside protection while recording healthy profits in market upswings.


\section{Conclusion}
An ultimate test for a risk methodology is to prove itself effective in sorting out risky and non-risky funds. We have tested two common risk measures against the newly proposed StressVaR in a test where we use the risk measure to allocate hedge funds within a hypothetical Fund-of-Funds. In all the tests the portfolio allocated using the StressVaR outperformed the market by about 6\% annually and the other two measure by about 3\% to 4\% annually at a similar volatility. 

The core advantage of the StressVaR is that it efficiently uses information external to the fund returns (the factors) but in a way that is directly relevant to the risk of the fund in question.  With this in mind one could be unimpressed by the superior performance of the StressVaR since it is based on much more information than the traditional VaR measures based on past performance. However,  the challenge of efficient and focused use of extra information in assessing the risk of a hedge fund should not be underestimated.

The main challenges in constructing a StressVaR are in a complete and high-quality factorset; in the reliable factor selection and ranking mechanism; and in the estimation of the dynamic relation between the hedge fund and a factor.   In addition the estimation of the density of factor returns, which is a crucial step in the StressVar estimate, can benefit from advanced econometric models such as the GARCH class as well as pure economic studies.  There are large potential improvements to the StressVaR and it will be interesting to follow future research in this area. With a methodology of this kind, the formerly rigid boundary between risk-management and asset allocation is arguably fading.


\bibliographystyle{unsrt}
\bibliography{iiz7}

\end{document}